\input lanlmac


\def\th{\theta}

\def\cob{\delta}
\def\ep{\epsilon}

\def\Tr{{\rm Tr}}

\def\hf{{1\over 2}}

\def\o{\over}

\def\til#1{\widetilde{#1}}

\def\si{\sigma}

\def\del{\partial}

\def\bra{\langle}
\def\ket{\rangle}
\def\lf{\left}
\def\ri{\right}
\def\riya{\rightarrow}

\def\La{\Lambda}
\def\h#1{\widehat{#1}}

\def\al{\alpha}

\def\rt#1{\sqrt{#1}}

\def\st{\star}

\def\sitarel#1#2{\mathrel{\mathop{\kern0pt #1}\limits_{#2}}}

\def\tx{\widetilde{x\,}}
\def\ttx{\widetilde{\widetilde{x\,}}}

\def\np#1#2#3{{ Nucl. Phys.} {\bf B#1} (#2) #3}

\def\pln#1#2#3{{Phys. Lett.} {\bf B#1} (#2) #3}

\def\pr#1#2#3{{ Phys. Rev.} {\bf D#1} (#2) #3}

\def\hpt#1{{\tt hep-th/#1}}


\lref\corn{
L. Cornalba and R. Schiappa,
``Matrix Theory Star Products from the Born-Infeld Action,''
\hpt{9907211}\semi
L. Cornalba,
`` D-brane Physics and Noncommutative Yang-Mills Theory,''
\hpt{9909081}.
}
\lref\Ishin{
N. Ishibashi,
``A Relation between Commutative and Noncommutative Descriptions 
of D-branes,'' \hpt{9909176}.
}

\lref\barbon{
E. Alvarez and J. L. F. Barb\'{o}n and J. Borlaf,
``T-duality for Open Strings,''
\np{479}{1996}{218}, \hpt{9603089}.
}
\lref\callan{
C. G. Callan, C. Lovelace, C. R. Nappi and S. A. Yost,
``Loop Corrections to Superstring Equations of Motion,''
\np{308}{1988}{221}. 
}

\lref\KO{
K. Okuyama,
``Path Integral Representation of the Map between Commutative and
Noncommutative Gauge Fields,''
JHEP  {\bf 0003} (2000) 016, \hpt{9910138}.
}

\lref\KatoKuro{
M. Kato and T. Kuroki,
``World Volume Noncommutativity versus Target Space Noncommutativity,''
JHEP {\bf 9903} (1999) 012, \hpt{9902004}. 
}
\lref\Hashimoto{
K. Hashimoto,
``Corrections to D-brane Action and Generalized Boundary State,''
Phys. Rev. {\bf D61}, 106002 (2000), \hpt{9909027}. 
}
\lref\Zhang{
Y. Zhang,
``A Note on the Path Integral Representation of the 
Boundary State of D-brane,''
\hpt{0005176}.
}

\lref\SW{
N. Seiberg and E. Witten,
``String Theory and Noncommutative Geometry,''
JHEP {\bf 9909} (1999) 032, \hpt{9908142}.
}

\lref\Ishibashietal{
N. Ishibashi, S. Iso, H. Kawai and Y. Kitazawa,
``Wilson Loops in Noncommutative Yang Mills,''
Nucl. Phys. {\bf B573} (2000) 573, \hpt{9910004}.
}
\lref\Kuroki{
T. Kuroki,
``Noncommutativities of D-branes and $\theta$-changing
Degrees of Freedom in D-brane Matrix Models,''
Phys. Lett. {\bf B481} (2000) 97, \hpt{0001011}.
}
\lref\Seiberg{
N. Seiberg,
``A Note on Background Independence in Noncommutative Gauge Theories, Matrix
Model and Tachyon Condensation,''
JHEP {\bf 0009} (2000) 003, \hpt{0008013}. 
}
\lref\BvsTh{
B. Pioline  and A. Schwarz,
``Morita equivalence and T-duality (or B versus $\Theta$),''
JHEP {\bf 9908} (1999) 021, \hpt{9908019}. 
}
\lref\Ryang{
S. Ryang,
``Open String and Morita Equivalence of the Dirac-Born-Infeld Action with
Modulus $\Phi$,'' \hpt{0003204}.
}

\lref\KT{
T. Kawano and T. Takahashi,
``Open String Field Theory on Noncommutative Space,'',
\hpt{9912274};
``Open-Closed String Field Theory in the Background B-Field,''
\hpt{0005080}. 
}
\lref\Sugino{
F. Sugino,
``Witten's Open String Field Theory in Constant B-Field Background,''
JHEP {\bf 0003} (2000) 017, \hpt{9912254}. 
}

\lref\chuho{
C.-S. Chu and P.-M. Ho,
``Noncommutative Open String and D-brane,''
\np{550}{1999}{151}, \hpt{9812219};
``Constrained Quantization of Open String in Background B Field 
and Noncommutative D-brane,'' 
Nucl. Phys. {\bf B568} (2000) 447, \hpt{9906192}.
}
\lref\jab{
F. Ardalan, H. Arfaei and M. M. Sheikh-Jabbari,
`` Noncommutative Geometry From Strings and Branes,''
JHEP {\bf 9902} (1999) 016, \hpt{9810072};
``Dirac Quantization of Open Strings and Noncommutativity in Branes,''
Nucl. Phys. {\bf B576} (2000) 578, \hpt{9906161}\semi
M. M. Sheikh-Jabbari,
``Open Strings in a B-field Background as Electric Dipoles,''
\pln{455}{1999}{129}, \hpt{9901080}.
}
\lref\scho{
V. Schomerus,
`` D-branes and Deformation Quantization,''
JHEP {\bf 9906} (1999) 030, \hpt{9903205}.
}
\lref\ACNY{
A. Abouelsaood, C. G. Callan, C. R. Nappi and S. A. Yost,
``Open Strings in Background Gauge Fields,''
\np{280}{1987}{599}.
}

\lref\Ishio{
N. Ishibashi,
``$p$-branes from $(p-2)$-branes in the Bosonic String Theory,''
\np{539}{1999}{107}, \hpt{9804163}.
}
\lref\branes{
T. Banks, N. Seiberg and S. Shenker,
``Branes from Matrices,''
\np{490}{1997}{91}, \hpt{9612157}.
}
\lref\BFFS{
T. Banks, W. Fischler, S. H. Shenker and L. Susskind,
``M Theory As A Matrix Model: A Conjecture,''
\pr{55}{1997}{5112}, \hpt{9610043}.
}
\lref\town{
P. K. Townsend,
``D-branes from M-branes,''
\pln{373}{1996}{68}, \hpt{9512062}.
}

\lref\NCLa{
M. M. Sheikh-Jabbari,
``On the Deformation of $\Lambda$-Symmetry in B-field Background,''
Phys. Lett. {\bf B477} (2000) 325, \hpt{9910258}.
}
\lref\Lerda{
P. Di Vecchia, M. Frau, A. Lerda and A. Liccardo,
``(F,Dp) bound states from the boundary state,''
Nucl. Phys. {\bf B565} (2000) 397, \hpt{9906214}.
}

\Title{             
                                             \vbox{\hbox{KEK-TH-715}
                                             \hbox{hep-th/0009215}}}
{\vbox{
\centerline{Boundary States in $B$-Field Background}
}}

\vskip .2in

\centerline{
                    Kazumi Okuyama
}

\vskip .2in

\centerline{\sl Theory Group, KEK, Tsukuba, Ibaraki 305-0801, Japan}
\centerline{\tt kazumi@post.kek.jp }

\vskip 3cm
\noindent

We consider the boundary states 
which describe  $D$-branes in a constant $B$-field background.
We show that the two-form field $\Phi$, 
which interpolates commutative and noncommutative descriptions of $D$-branes,
can be interpreted as the invariant field strength 
in the T-dual picture. We also show that the extended algebra parametrized by
$\th$ and $\Phi$ naturally appears as the commutation relations
of the original  and the T-dual coordinates.

\Date{September 2000}

\vfill
\vfill

\newsec{Introduction}
The short distance geometry seen by the extended objects
such as $D$-branes is one of the most interesting
problem in the string theory. 
The study of the $D$-brane worldvolume geometry is 
closely related to the study of geometry transverse to the branes
in view of the Matrix Theory construction of $D$-branes from the lower 
dimensional ones. 
In the perturbative open string theory,
the geometry on  $D$-branes can be probed by 
examining the OPE algebra of vertex  
operators of open strings ending on the branes. 
Therefore, the idea of the noncommutative geometry which identifies
the algebra and the geometry fits nicely with the open string description 
of $D$-branes.
From the equivalent description of $D$-branes in the dual channel,
i.e., the boundary state formalism,  
it is usually difficult to extract the information of
the worldvolume geometry of branes.
However, in some cases the boundary state formalism is more powerful than 
the open string approach to analyze the geometry on $D$-branes.

In the background of the constant metric $g$ and 
the NSNS two-form field $B$, 
the algebra of open string vertex operators \refs{\ACNY,\chuho,\jab,\scho}
and the boundary state formalism \refs{\Ishio,\Ishin,\KatoKuro,\KO}
lead to the same worldvolume geometry of $D$-branes, 
which is characterized by the relation
\eqn\comxxth{
[x^i,x^j]=i\th^{ij}.
} 
In \SW, Seiberg and Witten argued that the theory on $D$-branes in 
a $B$-field background can be described by either commutative or
noncommutative Yang-Mills theories. 
They also proposed that there is a family of descriptions 
parametrized by the two-form field $\Phi$
which interpolates
these two descriptions.
The closed string background $g,B$ and the open string parameters $G,\th,\Phi$
are related by
\eqn\Phidef{
{1\o g+B}=\th+{1\o G+\Phi}.
}
In this paper, we use the convention $2\pi\al'=1$. 

In \Seiberg, the extended algebra 
which includes \comxxth\ as a subalgebra was proposed. It is defined by 
\eqn\Seialg{
[x^i,x^j]=i\th^{ij},\quad
[\del_i,x^j]=\cob_i^j,\quad
[\del_i,\del_j]=-i\Phi_{ij}.
}
From this algebra, we can construct the variables which make it manifest
that the theory on $D$-branes in a $B$-field background 
depends only on the combination $\h{F}+\Phi$.
In this paper, we study the string theory origin of the variable $\del_i$
and the algebra \Seialg\ from the boundary state formalism.  
We show that $\del_i$ is obtained by transforming
$x^i$ by the $\La$-symmetry in the T-dual picture.

This paper is organized as follows:
In section 2, we review the construction of boundary states in a
$B$-field background. In section 3, we present three descriptions of 
boundary states in a constant $B$-field background: 
the path integral representation, 
the operator formalism and the matrix model representation.
In section 4, we consider the action of T-duality on boundary states.
In section 5, we show that $\Phi$ is related to the 
flux invariant under the $\La$-symmetry in the T-dual picture.
In section 6, we argue that the algebra \Seialg\ naturally appears by 
performing the $\La$-transformation in the T-dual picture.
Section 7 is devoted to  discussions.

\newsec{Boundary States in $B$ Field Background}
To make this paper self-contained, we start with a review of the
construction of boundary states in a $B$-filed background in the bosonic
string theory.\foot{For the boundary state in a constant 
$B$-field background in 
superstring theory, see e.g. \Lerda.}

\subsec{Coherent State}

The boundary state is a state in the Hilbert space 
of a closed string satisfying some boundary conditions. 
The worldsheet action of the closed string in the background of constant 
metric $g_{ij}$ and NSNS 2-form filed $b_{ij}$
is given by
\eqn\stract{
S_{{\rm closed}}=
\int d\tau\int_0^{2\pi} d\si \lf[\hf g_{ij}(\del_{\tau}X^i\del_{\tau}X^j
-\del_{\si}X^i\del_{\si}X^j)-b_{ij}\del_{\tau}X^i\del_{\si}X^j\ri].
}
The conjugate momentum of $X^i$ is defined by
\eqn\PvsB{
P_i=g_{ij}\del_{\tau}X^j-b_{ij}\del_{\si}X^j,
}
which satisfies $[X^i(\si),P_j(\si')]=i\cob^i_j\cob(\si-\si')$.
We take the boundary of the string to be at $\tau=0$ 
and parametrized by $\si$. 
In the open string picture,
the boundary states correspond to the boundary conditions at $\si=0$  
of the open string  with the action
\eqn\openS{
S_{{\rm open}}=
\int d\tau\int_0^{\pi}d\si\lf[\hf g_{ij}(\del_{\tau}X^i\del_{\tau}X^j
-\del_{\si}X^i\del_{\si}X^j)+b_{ij}\del_{\tau}X^i\del_{\si}X^j\ri].
}
The difference of the sign in front of $b$ in \stract\ and \openS\ comes from 
the fact that the orientation of the worldsheet is reversed
when we make the identification $(\tau_{{\rm closed}},\si_{{\rm closed}})
=(\si_{{\rm open}},\tau_{{\rm open}})$.
    
To construct the general boundary states, it is useful to introduce the
coherent state $|x\ket$ \callan\ which is defined by
\eqn\cohdef{
X^i(\si)|x\ket=x^i(\si)|x\ket.
}
For example, using $|x\ket$ 
the Dirichlet state $|D\ket$ and the Neumann state
$|N\ket$ is written as
\eqn\NDinx{\eqalign{
|D\ket&=|x={\rm const.}\ket, \cr
|N\ket&=\int[dx]|x\ket=\int[dx]\exp\left(-i\int d\si P_i(\si)x^i(\si)\right)
|x=0\ket,
}}
where $\int[dx]$ denotes the path integral over the boundary loop $x(\si)$.
We can easily see that these states satisfy the boundary conditions 
\eqn\defND{
P_i(\si)|N\ket=0,\quad \del_{\si}X^i(\si)|D\ket=0.
}

\subsec{Gauge Fields and $\La$-Symmetry}
In the closed string picture, gauge fields can be incorporated in
boundary states as
Wilson loops:
\eqn\Bact{\eqalign{
|B\ket&=\int[dx]\exp\lf(i\int d\si A_i(x)\del_{\si}x^i\ri)|x\ket \cr
&=\int[dx]\exp\lf(i\int d\si \Big(A_i(x)\del_{\si}x^i-P_ix^i\Big)\ri)|x=0\ket.
}}
This state formally satisfies the mixed boundary condition
\eqn\mixedB{
(P_i-F_{ij}(X)\del_{\si}X^j)|B\ket=0.
}
For the case of non-constant $F_{ij}$, this expression contains
divergence. In this paper, we only consider the case of constant $F_{ij}$. 
From the definition of $P_i$ \PvsB, this boundary condition depends only
on the combination $b+F$.
This is a consequence of the so-called $\La$-symmetry.
This symmetry is generated by
\eqn\Ulambda{
U_{\La}=\exp\lf(i\int\La_i(X)\del_{\si}X^i\ri).
}
Under the $\La$-symmetry, $P_i$ transforms as
\eqn\LAU{
U_{\La}P_iU_{\La}^{-1}
=P_i-({\del_i}\La_j-\del_j\La_i)\del_{\si}X^j,
}
or
\eqn\Btobpla{
U_{\La}P_i(b)U_{\La}^{-1}=P_i(b+d\La).
}
Using $U_{\La}|x=0\ket=|x=0\ket$, we can rewrite the boundary state as
\eqn\AtoAla{\eqalign{
|B\ket&=\int[dx]\exp\lf(i\int d\si \Big(A_i(x)\del_{\si}x^i
-P_i(b)x^i\Big)\ri)|x=0\ket \cr
&=\int[dx]e^{i\int d\si A_i\del_{\si}x^i
}U_{\La}^{-1}U_{\La}e^{-i\int d\si P_i(b)x^i}U_{\La}^{-1}
|x=0\ket\cr
&=\int[dx]\exp\lf(i\int d\si \Big(\big(A_i(x)-\La_i(x)\big)\del_{\si}x^i
-P_i(b+d\La)x^i\Big)\ri)|x=0\ket.
}}
Therefore, the boundary state has the symmetry\foot{See \NCLa\ for
the $\La$-symmetry in the noncommutative description of $D$-brane.}
\eqn\Lasym{
A\riya A-\La,\quad b\riya b+d\La
}
which leaves the combination ${\cal F}=b+F$ invariant.

\newsec{Three Representations of Boundary States}
The boundary state for the $D$-brane with constant field strength $f$ on it
is written as
\eqn\constF{
|B\ket=\int[dx]\exp\lf(i\int d\si \Big(
\hf x^if_{ij}\del_{\si}x^i-P_ix^i\Big)\ri)|x=0\ket.
}
Note that the information of the background $g$ and $b$ are contained in $P_i$.
In this section, we see that this state can be 
written in three equivalent ways.

\subsec{Path Integral Representation}
The first is the original path integral representation
\constF.
From the kinetic term for $x^i$, we can read off the propagator of $x^i$ to be
\eqn\propx{
\bra x^i(\si)x^j(\si')\ket=[(-if\del_{\si})^{-1}]^{ij}
={i\o2}\th^{ij}\ep(\si-\si'),
}
where $\ep(\si)$ is the sign function and $\th$ is defined by
\eqn\finvth{
\th=f^{-1}.
}
By the point splitting regularization, the equal time commutators of
$x^i$ turn out to be
\eqn\commx{
[x^i(\si),x^j(\si)]=i\th^{ij}.
}
This is the closed string description of the noncommutativity
of the boundary coordinates of the open string.

\subsec{Operator Representation}
The second representation of \constF\ is the operator representation.
By performing the Gaussian integral over $x$ in \constF, we  
obtain
\eqn\oprep{
|B\ket=V_{\th}|x=0\ket
}
where $V_{\th}$ is given by
\eqn\KTop{
V_{\th}=
\exp\left(-{i\o4}\int d\si d\si'P_i(\si)\th^{ij}\ep(\si-\si')P_j(\si')\right).
}
Note that this unitary operator $V_{\th}$ has the same form as the one 
introduced in 
\KT\ to construct the 3-string
vertex of the open string field theory in the presence of a background
$B$-field. (See also \Sugino.) Therefore,  we call this operator
the KT operator 
in the following.
The action of KT operator on the closed string coordinate $X^i$ 
is found to be
\eqn\VthonX{
V_{\th}X^i(\si)V_{\th}^{-1}=X^i(\si)
-\hf\int d\si'\ep(\si-\si')\th^{ij}P_j(\si'),
}
and its derivative $\del_{\si}X^i$ transforms as
\eqn\BCchange{
V_{\th}\del_{\si}X^iV_{\th}^{-1}=\del_{\si}X^i-\th^{ij}P_j.
}
Using this relation and $\del_{\si}X^i|x=0\ket=0$,
we can check that $|B\ket$ in \oprep\  satisfies the mixed boundary condition
\eqn\mixthB{
(\del_{\si}X^i-\th^{ij}P_j)|B\ket=0
}
which is the same as \mixedB\ with $F_{ij}(X)=f_{ij}$. 

\subsec{Matrix Representation}
The third one is the matrix representation. Since the path integral in 
\constF\ is taken 
over the variable $x^i(\si)$
with periodic boundary condition, $|B\ket$ can be written as
a trace over the Hilbert space on which $x^i$ satisfies the commutation
relation \comxxth. Using the star product 
\eqn\defst{
f\st g=f\exp\left({{i\o2}\th^{ij}\overleftarrow{\partial_i}
\overrightarrow{\partial_j}}\right)g,
}
$|B\ket$ is written as
\eqn\matrep{
|B\ket=\Tr \,P\exp_{\st}\left(-i\int d\si P_i(\si)x^i\right)|x=0\ket 
}
where the product of $x^i$ is taken by $\st$ and 
$P$ denotes the path ordering.
Since $|x=0\ket$ satisfies the Dirichlet boundary condition
for all directions, it represents a $D$-instanton at the origin.
Therefore,
the state $|B\ket$ in the form of \matrep\ can be interpreted that the 
matrix coordinates $\phi^i$ of infinitely many $D$-instantons 
have their value $\phi^i=x^i$ \refs{\Ishio,\town,\BFFS,\branes}. 
Note that this picture can  also be applied to the construction
of $p$-brane from infinitely many $p'$-brane with $p=p'+{\rm rank}f$.

The more generic configuration of $D$-instantons around this classical 
solution can be represented
in the matrix representation as
\eqn\genmat{
|B\ket=\Tr \,P\exp_{\st}\left(-i\int d\si P_i(\si)\phi^i(x)\right)|x=0\ket, 
} 
and in the path integral representation as
\eqn\pathphi{
|B\ket=\int[dx]\exp\left(i\int d\si\Big(\hf x^if_{ij}\del_{\si}x^j
-P_i\phi^i(x)\Big)\right)|x=0\ket.
}
The noncommutative gauge field $\h{A}_i$
appears as the fluctuation
around the classical solution $\phi^i=x^i$ \refs{\Ishin,\corn}:
\eqn\hAandphi{
\phi^i=x^i+\th^{ij}\h{A}_j(x).
}
These matrices satisfy
\eqn\matcom{
[\phi^i,\phi^j]=-i(\th\h{F}\th-\th)^{ij}=-i
\Big\{\th(\h{F}-\th^{-1})\th\Big\}^{ij}.
}

\newsec{T-duality of Boundary States}
In this section, we consider the action of T-duality on boundary states.
T-duality for closed string coordinates is given by
\eqn\TdualX{\eqalign{
\del_{\si}\til{X_i}&=g_{ij}\del_{\tau}X^j-b_{ij}\del_{\si}X^j=P_i,\cr
\del_{\tau}\til{X_i}&=g_{ij}\del_{\si}X^j-b_{ij}\del_{\tau}X^j.
}}
The closed string background for the T-dual variable are related to the
original background $g$ and $b$ by
\eqn\gbtotgtb{
\til{g}+\til{b}={1\o g+b}. 
}

The Dirichlet boundary state in the original variables 
is written as a Neumann boundary state in
the T-dual picture
\eqn\DtoN{
|x=0\ket=\int[d\til{x}]|\til{x}\ket.
}
From this, the relation between 
the coherent state for the original and the T-dual coordinates
is found to be \barbon
\eqn\xkettxket{
|x\ket=\exp\lf(-i\int d\si P_ix^i\ri)|x=0\ket=\int[d\til{x}]
\exp\lf(-i\int d\si \del_{\si}\til{x_i}x^i\ri)|\til{x}\ket.
}

\subsec{KT Operator and $\La$-Symmetry}
In the T-dual picture,
the KT operator has a simple meaning.
By substituting $P_i=\del_{\si}\til{X_i}$
in $V_{\th}$, it becomes 
\eqn\Vthindual{
V_{\th}=\exp\left(i\int d\si\hf\til{X_i}\th^{ij}\del_{\si}\til{X_j}\right).
}
This shows that the KT operator $V_{\th}$ shifts the field strength
by an amount $\th$ in the T-dual picture.
In other words, $V_{\th}$ can be written as a combination of 
$\La$-symmetry and T-duality:
\eqn\VtoU{
V_{\th}=TU_{d^{-1}\th}T
}
where $d^{-1}\th$ is a one-form defined by
\eqn\defdinvth{
d^{-1}\th=\hf\til{X_i}\th^{ij}d\til{X_j}.
}

\newsec{$\Phi$ as T-dual Flux}
In \SW, the two-form field $\Phi$ was introduced to 
interpolate the commutative and noncommutative descriptions
of $D$-branes in a $B$-field background.
In \Ishibashietal, 
it was argued that this degree of freedom corresponds to the 
freedom of splitting the flux invariant under $\La$-symmetry into 
the gauge field strength on $D$-brane $f$ and the NSNS 2-form field $b$ 
We denote this $\La$-invariant flux by $B$, i.e.,
\eqn\Binomb{
B=b+f.
}

In \Ishibashietal, it was shown that
the open string metric $G$ and the two-form $\Phi$ 
can be read off from the effective action for 
infinitely many $D$-instantons. (See also \refs{\BvsTh,\Kuroki,\Ryang} for 
the interpretation of $\Phi$.) 
The argument in \Ishibashietal\ is as follows:
The Lagrangian for $D$-instantons is proportional to
\eqn\Laggb{
\rt{\det(\til{g}^{ij}+\til{b}^{ij}-i[\phi^i,\phi^j])}.
}
Using \matcom, this can be written as
\eqn\BIhF{
\det\th\,\rt{\det(G+\Phi+\h{F})}
}
with
\eqn\GPhi{\eqalign{
G&=f^T\til{g}f=-f\til{g}f,\cr
\Phi&=f^T(\til{b}+\th)f=-f(\til{b}+\th)f.
}}
From \gbtotgtb, one can  show that $G$ and $\Phi$ of this form 
satisfy the relation \Phidef.

We can derive the same relation by using the T-duality for the boundary state 
which clarifies the meaning of $\Phi$.
In the T-dual picture, the boundary state \constF\ becomes
\eqn\Bounintilx{\eqalign{
|B\ket&=\int[dx]\exp\lf(i\int d\si\Big(\hf x^if_{ij}\del_{\si}x^j
-P_ix^i\Big)\ri)|x=0\ket \cr
&=\int[dxd\tx]\exp\lf(i\int d\si\Big(\hf x^if_{ij}\del_{\si}x^j
-x^i\del_{\si}\tx_i-\til{P}^i\tx_i\Big)\ri)|\til{x}=0\ket \cr 
&=\int[d\tx]\exp\lf(i\int d\si\Big(\hf \til{x_i}\th^{ij}\del_{\si}\til{x_j}
-\til{P}^i\tx_i\Big)\ri)|\til{x}=0\ket. 
}}
From this form of boundary state,
$\th$ can be interpreted as the gauge field strength in the T-dual 
picture.
By adding the NSNS two-form field $\til{b}$ to it,
the $\La$-invariant flux $\til{{\cal F}}$ in the T-dual picture becomes
\eqn\tilF{
\til{{\cal F}}=\til{b}+\th.
}  

Since the coupling to the noncommutative gauge field has the form
\eqn\hAtotx{
-i\int d\si P_i\th^{ij}\h{A}_j=i\int d\si \h{A}_i\th^{ij}\del_{\si}\til{X_j},
}
we should raise the index of $\til{X_i}$ by
\eqn\tXsupi{
\til{X^i}=\th^{ij}\til{X_j}
}
so that the coupling \hAtotx\ has the canonical form. 
By the change of variables from $\til{X_i}$ to
$\til{X^i}$, the metric $\til{g}^{ij}$ and the invariant flux
$\til{{\cal F}}^{ij}$ are transformed to $G_{ij}$ and $\Phi_{ij}$ in \GPhi.
Therefore, $\Phi$ can be interpreted as the invariant flux $\til{{\cal F}}$ 
written in the coordinate $\til{X^i}=\th^{ij}\til{X_j}$:
\eqn\PhiandtF{
\Phi=f^T\til{\cal F}f.
}
Note that $\Phi$ itself is not $\La$-invariant while $\til{{\cal F}}$ is,
since $f$ is not invariant under the $\La$-symmetry.

\newsec{Origin of Extended Algebra}
In this section, we consider the origin of the algebra
\Seialg\ from the viewpoint of the  boundary state formalism.
In the following discussion, it is convenient
to introduce the variable $\til{\til{x^i}}$ by
\eqn\deltottx{
\til{\til{x^i}}=i\th^{ij}\del_j.
}
In terms of $x^i$ and $\til{\til{x^i}}$, the algebra \Seialg\ is written as
\eqn\tilFalg{
[x^i,x^j]=i\th^{ij},\quad [\til{\til{x^i}},x^j]=i\th^{ij},\quad 
[\til{\til{x^i}},\til{\til{x^j}}]=i\til{{\cal F}}^{ij}.
}

\subsec{Algebra from Boundary State Formalism}
Using $x^i$ and $\til{\til{x^i}}$, we can 
write the Lagrangian \Laggb\ as
\eqn\Lagintphi{
\rt{\det(\til{g}^{ij}+\til{b}^{ij}-i[\phi^i,\phi^j])}
=\rt{\det(\til{g}^{ij}-i[\til{\til{\phi^i}},\til{\til{\phi^j}}])}
}
where $\til{\til{\phi^i}}$ is given by
\eqn\ttphi{
\til{\til{\phi^i}}=\til{\til{x^i}}+\th^{ij}\h{A}_j(x).
}
These variables satisfy the relation
\eqn\conttphi{
[\til{\til{\phi^i}},\til{\til{\phi^j}}]
=-i(\th\h{F}\th-\til{{\cal F}})^{ij}
=-i\Big\{\th(\h{F}+\Phi)\th\Big\}^{ij}.
}
Changing from the variable $\phi^i$ to $\til{\til{\phi^i}}$
is equivalent to set $\til{b}=0$ by the $\La$-symmetry.
Since this $\La$-symmetry is performed in the T-dual picture,
we can guess that $\til{\til{x^i}}$ is given by
\eqn\ttxvsx{
\til{\til{x^i}}=TU_{-d^{-1}\til{b}}T(x^i).
}

In the following, using the path integral representation of boundary states
we show that the variables defined by this relation satisfy 
the algebra \tilFalg.
Following \ttxvsx, we first consider the T-dual of
the state \pathphi:
\eqn\BinoneT{\eqalign{
|B\ket&=\int[dx]\exp\lf(i\int d\si\Big(\hf x^if_{ij}\del_{\si}x^j-
P_i(x^i+\th^{ij}\h{A}_j)\Big)\ri)|x=0\ket \cr
&=\int[dxd\tx]\exp\lf(i\int d\si\Big(\hf x^if_{ij}\del_{\si}x^j
-\del_{\si}\tx_i(x^i+\th^{ij}\h{A}_j)
-\til{P}^i\tx_i\Big)\ri)|\til{x}=0\ket.
}}
Next we perform the 
$\La$-transformation in order to set $\til{b}=0$ in $\til{P^i}$:
\eqn\Blatbzero{\eqalign{
|B\ket=\int[dxd\tx]\exp&\lf(i\int d\si\Big(\hf x^if_{ij}\del_{\si}x^j
-\del_{\si}\tx_i(x^i+\th^{ij}\h{A}_j)\ri. \cr
&\qquad \lf.+\hf\tx_i\til{b}^{ij}\del_{\si}\tx_j
-\til{P}^i(\til{b}=0)\tx_i\Big)\ri)|\til{x}=0\ket.
}}
Taking the T-duality one more time, the boundary state becomes
\eqn\Bxtxttx{\eqalign{
|B\ket=\int[dxd\tx d\ttx]\exp &\lf(i\int d\si\Big(\hf x^if_{ij}\del_{\si}x^j
-\del_{\si}\tx_i(x^i+\th^{ij}\h{A}_j) \ri.
\cr 
&\qquad\qquad
\lf.+\hf\tx_i\til{b}^{ij}\del_{\si}\tx_j
-\tx_i\del_{\si}\til{\til{x^i}}-\til{\til{P}}_i\til{\til{x^i}}\Big)\ri)
|\til{\til{x}}=0\ket.
}}
By integrating out the variable $\til{x^i}$, we arrive at the 
final expression:
\eqn\BintwoT{\eqalign{
|B\ket
&=\int[dxd\ttx]\exp\lf(i\int d\si\Big(\hf x^if_{ij}\del_{\si}x^j
+\hf(\til{\til{x^i}}-x^i-\th^{ik}\h{A}_k)(\til{b}^{-1})_{ij}\del_{\si}
(\til{\til{x^j}}-x^j-\th^{jl}\h{A}_l)\ri.\cr
&\qquad\qquad\qquad\qquad\qquad\qquad\qquad\qquad
-\til{\til{P_i}}\til{\til{x^i}}\Big)\bigg)|\til{\til{x}}=0\ket\cr
&=\int[dxd\ttx]\exp\lf(i\int d\si\Big(\hf x^if_{ij}\del_{\si}x^j
+\hf(\til{\til{x^i}}-x^i)(\til{b}^{-1})_{ij}\del_{\si}
(\til{\til{x^j}}-x^j)\ri.\cr
&\qquad\qquad\qquad\qquad\qquad\qquad\qquad\qquad
-\til{\til{P_i}}(\til{\til{x^i}}+\th^{ij}\h{A}_j)
\Big)\bigg)|\til{\til{x}}=0\ket.
}}
In the last step, we shifted the integration variable $\til{\til{x^i}}$ 
to $\til{\til{x^i}}+\th^{ij}\h{A}_j$. 
From this expression of $|B\ket$, 
the commutation relations of $x^i$ and $\til{\til{x^i}}$ are found to be
\eqn\comtxx{
[x^i,x^j]=i\th^{ij},\quad
[\til{\til{x^i}}-x^i,x^j]=0,\quad
[\til{\til{x^i}}-x^i,\til{\til{x^j}}-x^j]=i\til{b}^{ij}.
}
We can see that these relations are equivalent to \tilFalg.
In this form of $|B\ket$,  $\til{\til{\phi^i}}$ in \ttphi\
naturally appear as the matrix coordinates 
of $D$-instantons in the $\til{\til{x^i}}$ frame.

\subsec{Relation between $x^i$ and $\til{\til{x^i}}$}
In this subsection, we show that the variable $\til{\til{x^i}}$ 
appeared in the boundary state
\BintwoT\ actually satisfies the relation \ttxvsx.
In the operator formalism, this relation can be rephrased as 
$\til{\til{X^i}}$ being the transformation of $X^i$ 
by the KT operator:
\eqn\KTtrfX{
\til{\til{X^i}}(\si)=V_{\til{b}}^{-1}X^i(\si)V_{\til{b}} 
=X^i(\si)+{1\o2}\int d\si'\ep(\si-\si')\til{b}^{ij}P_j(\si').
}

To derive this relation  from the boundary state formalism,
let us consider the identity
\eqn\idzero{\eqalign{
0=\int[dxd\ttx]{\cob\o\cob\til{\til{x^i}}(\si)}
\exp&\lf(i\int\hf x^if_{ij}\del_{\si}x^j
+\hf(\til{\til{x^i}}-x^i)(\til{b}^{-1})_{ij}\del_{\si}
(\til{\til{x^j}}-x^j)\ri.\cr
&\qquad\qquad\qquad
-\til{\til{P_i}}(\til{\til{x^i}}+\th^{ij}\h{A}_j)\Big)|\til{\til{x}}=0\ket.
}}
From this identity we obtain
\eqn\idonB{
\Big[\del_{\si}(\til{\til{X^i}}-X^i)-\til{b}^{ij}\til{\til{P_j}}\Big]|B\ket=0,
}
which is equivalent to \KTtrfX\ since
\eqn\PtottP{
P_i=\del_{\si}\til{X_i}=\til{\til{P_i}}.
}
From the boundary condition for $X$ \mixthB, the boundary condition for
$\til{\til{X}}$ becomes
\eqn\ttXBC{
(\del_{\si}\til{\til{X^i}}-\til{{\cal F}}^{ij}P_j)|B\ket=0.
}
Note that 
$\til{\til{x^i}}$ is identical to $x^i$ when $\til{b}=0$, or $b=0$. As 
was pointed out in \Seiberg,
this case corresponds to $\Phi=-\th^{-1}$.

\newsec{Discussions}
In this paper, we derived the extended algebra \Seialg\ parametrized by
$\th$ and $\Phi$ from the boundary state formalism.
We identified the extra variable $\del_i$ as 
the KT transform of the original variable $x^i$, i.e., the $\La$-transform
of $x^i$ in the T-dual picture. This algebra may shed light on the
background independent description of the dynamics of $D$-branes.

Although we can construct the extended phase space $(x^i,\til{\til{x^i}})$
in the closed string picture, its interpretation in the open 
string channel is not clear. However, we believe that
this extended algebra can be constructed by the variables 
in the open string theory. 

We comment on the relation between our interpretation of $\Phi$ and the
gauge fixing of worldvolume diffeomorphism on $Dp$-branes 
\refs{\Ishio,\Ishin,\corn}.
Before gauge fixing, the dynamical fields on a $Dp$-brane are the gauge
field $A_i$ and the scalar fields $\phi^i$.  After fixing the gauge field to 
have a constant field strength  $A=d^{-1}f$, the fluctuation of $\phi^i$
around the static gauge configuration can be identified with the 
noncommutative gauge field. Then the residual diffeomorphism ${\rm Diff}_f$ 
which preserves $f$ corresponds to the noncommutative gauge symmetry.
On the other hand, when we fix $\phi^i$ to the static gauge configuration, 
the fluctuation of $A_i$ around $d^{-1}f$ becomes the commutative gauge     
field. The author of \Kuroki\ argued that 
the diffeomorphism which changes the value of $f$,
${\rm Diff}/{\rm Diff}_f$, 
corresponds to the degree of freedom for $\Phi$.

In our discussion of $\Phi$, we always use $A=d^{-1}f$ gauge, so 
we are considering the noncommutative picture.
Note that the commutative picture is singular in 
our formalism since $\th=0$ corresponds to $f=\th^{-1}=\infty$.  
While preserving the property that the field strength  is constant,
we can change the value of  $f$ by using the $\La$-symmetry instead of 
the diffeomorphism.
In \Ishibashietal, this degree of freedom is identified with 
$\Phi$. Using this interpretation of $\Phi$, we explicitly performed the 
$\La$-symmetry on the boundary state and identified $\Phi$ within
this formalism.

\vskip 12mm
\centerline{{\bf Acknowledgments}}
I would like to thank T. Asakawa and N. Ishibashi  
for useful comments and discussions. 
I am also grateful to the organizers
of  ``Summer Institute 2000'' at Yamanashi, Japan, 
where a part of this work has been carried out.
This work was supported in part by JSPS Research Fellowships for Young
Scientists. 

\listrefs

\end